\documentclass{aa}
\usepackage{graphicx}
\usepackage{epsfig}

\begin{document}

\title{Preferred sunspot longitudes: Non-axisymmetry and differential rotation}

\offprints{I.G. Usoskin}

\author{I.G. Usoskin \inst{1}, S.V. Berdyugina\inst{2,3}
\and J. Poutanen\inst{2}}

\institute{Sodankyl\"a Geophysical Observatory (Oulu unit),
    FIN-90014 University of Oulu, Finland (ilya.usoskin@oulu.fi)
\and
    Astronomy Division, P.O. Box 3000,
         FIN-90014 University of Oulu, Finland (juri.poutanen@oulu.fi)
\and
         Institut f\"ur Astronomie, ETHZ, 8092 Z\"urich,
         Switzerland (sveta@astro.phys.ethz.ch)
}

\date{Received $<$date$>$/ accepted $<$date$>$}

\authorrunning{Usoskin, Berdyugina \& Poutanen}
\titlerunning{Non-axisymmetry in sunspot distribution}

\abstract {As recently found, the distribution of sunspots is non-axisymmetric and
 spot group formation implies the existence of two persistent active longitudes
 separated by 180\degr.
Here we quantitatively study the non-axisymmetry of sunspot occurrence.
In a dynamic reference frame inferred from the differential
 rotation law, the raw sunspot data show a clear clustering around the
 persistent active longitudes.
The differential rotation describing the dynamic frame is quantified in
 terms of the equatorial angular velocity and the differential rotation rate,
 which appear to be significantly different from those for individual sunspots.
This implies that the active longitudes are not linked to the depth of
 sunspot anchoring. In order to quantify the observed effect,
we introduce a measure of the non-axisymmetry of the sunspot
distribution. The non-axisymmetric component is found to be
highly significant, and the ratio
 of its strength to that of the axisymmetric one is roughly 1:10.
This  provides additional constraints for solar dynamo models.
 }

\maketitle


\section{Introduction}


The question whether sunspots appear randomly in longitudes has been
 a long-standing issue since the early 20th century.
Although the existence of preferred longitudes of sunspot
 formation (active longitudes) has been suggested long ago,
 the question of their persistency
 was still a subject of ongoing debates (e.g., \cite{chi32} 1932; \cite{lope61}
  1961; \cite{balt83} 1983; \cite{viti86} 1986; \cite{mord04} 2004).
A novel analysis of sunspot group data for the past 120 years
 revealed the existence of two persistent active
 longitudes separated by 180\degr\ (\cite{bu03} 2003, BU03 henceforth).
In BU03 we have shown, using different filtering techniques, that the
 active longitudes are {\it persistent} on a century time scale.
An important conclusion of our previous work is that the active longitudes
 are not fixed in any reference frame (e.g., in the Carrington system), but
 {\it continuously migrate} in longitude  with a variable rate. Their migration is
 defined by changes of the {\it mean latitude} of the sunspot formation and
 the {\it differential rotation}.
Neglecting this migration results in smearing of the active longitude
 pattern on time scales of more than one solar cycle.
 In contrast with the findings of BU03,
most previous researchers assumed rotation of the
active longitudes  with a constant rate, which explains the
 diversity and contradictions of the previously published results.

The solar active longitudes and their behaviour are
 very similar to stellar activity phenomena, including the
flip-flop cycle detected in binaries and solar-type stars
(\cite{bt98} 1998; \cite{b04} 2004; \cite{bj05} 2005).
On the Sun the flip-flop phenomenon is observed as the alternation
 of the major spot activity between the opposite longitudes
 with a 3.7 year cycle (BU03).
The similarity between the sunspot distribution and the activity
 patterns on cool active stars implies that non-axisymmetry is
 a fundamental feature of the solar and stellar dynamo mechanisms.

The persistent migrating active longitudes imply the
 existence of a non-axisymmetric component in the solar
 dynamo and provide new observational constraints
 for current solar dynamo models.
Therefore, it is important to quantify this effect.
As found by BU03, the migration of the active longitudes is defined
 by the differential rotation and mean latitude of sunspot formation.
In the present paper we fit this model
to raw sunspot data and determine the differential rotation of the active
longitudes. Based on that, we introduce a dynamic reference frame
and investigate the distribution of the sunspot area in this new frame.
This allows us to account for the migration of the active longitudes.
We reveal a double-peaked longitude distribution of the spot area
for all sunspots without any filtering and also for the single,
 strongest spot group observed in each Carrington rotation.
More impressively, such a distribution is also found for the
single, strongest spot group observed in each Carrington rotation.
Finally, we introduce a measure of the non-axisymmetry of the
sunspot distribution and estimate a relative strength of the
axisymmetric and non-axisymmetric components. The
differential rotation law obtained and the measure of the
non-axisymmetry can be used to constrain the corresponding dynamo
models. Our new analysis confirms the previous conclusions  by
BU03 on a new basis
 and dispels the doubts expressed by \cite{pelt04} (2005) that the active
longitude separation is an artefact of the data processing.

In this paper we analyse sunspot group data for the past 120 years.
We use daily data on sunspot group locations and areas collected at the Royal
 Greenwich Observatory, the US Air Force and the National Oceanic and
 Atmospheric Administration for the years 1878--1996, covering 11 full
 solar cycles.
Here we are primarily interested in the sunspot appearance rather than
 in their evolution.
Accordingly, each spot was included into the analysis only once when it was first
 mentioned in the database  (either on the day of its birth or when it appeared
at the East limb), all later records of the spot were ignored.
Therefore, in contrast to earlier studies, we analyze only the sunspot emergence.
Because of the known asymmetry between the Northern and Southern hemispheres, we
  investigate them separately.
About 40,000 sunspots occurred during about 1600
 Carrington rotations have been considered in each hemisphere.


\section{A dynamic reference frame}


As suggested in BU03, the migration of the active longitudes
 on the Sun is defined by the differential rotation and by changes of the
 mean spot latitude.
A standard model of the differential rotation on the Sun relates
an angular velocity $\Omega$ with a helio-latitude $\psi$ as follows:
\begin{equation}
\Omega=\Omega_{\rm 0}-B\,\sin^2\psi-C\,\sin^4\psi,
\label{Eq:Omega}
\end{equation}
where $\Omega_{\rm 0}$ is the equatorial angular velocity, while $B$ and $C$
describe the differential rotation rate.
Here we do not use the higher order term $C\sin^4\psi$, because it influences
mostly higher latitudes where sunspots are not observed.
Throughout the paper, we use the time step equal to the Carrington period,
 and Eq.~(\ref{Eq:Omega}) takes the form
\begin{equation}
\Omega_i=\Omega_{\rm 0}-B\,\sin^2\langle\psi\rangle_i,
\label{Eq:Omegai}
\end{equation}
where index $i$ denotes the $i$-th Carrington rotation, and $\langle\psi\rangle_i$
 is the area weighted average latitude of sunspots during this Carrington
 rotation.
Based on that, we introduce a new reference frame which describes the
 longitudinal migration of active regions with respect to the Carrington frame
 due to differential rotation:
\begin{equation}
\Lambda_i=\Lambda_0+T_{\rm C}\sum_{j=N_0}^{i}\left(\Omega_{\rm C}-\Omega_j\right),
\label{Eq:migr}
\end{equation}
where $\Lambda_i$ is the expected active longitude in the $i$-th
 Carrington rotation, $\Lambda_0$ is the location of the active longitude
 in the $N_0$-th Carrington rotation,
$T_{\rm C}\!=\!25.38$ days is the sidereal Carrington period
and $\Omega_{\rm C}\!=\!360^\circ/T_{\rm C}$. In our model  $\Omega_0$
and $B$ are the parameters (Eq.~\ref{Eq:Omegai}) that need to be
determined from observations, while
 $\Lambda_0$ is just a constant defining a shift in longitude of the
new reference frame with respect to the Carrington system
 in the $N_0$-th Carrington rotation.

\begin{figure}
\centerline{\epsfig{file=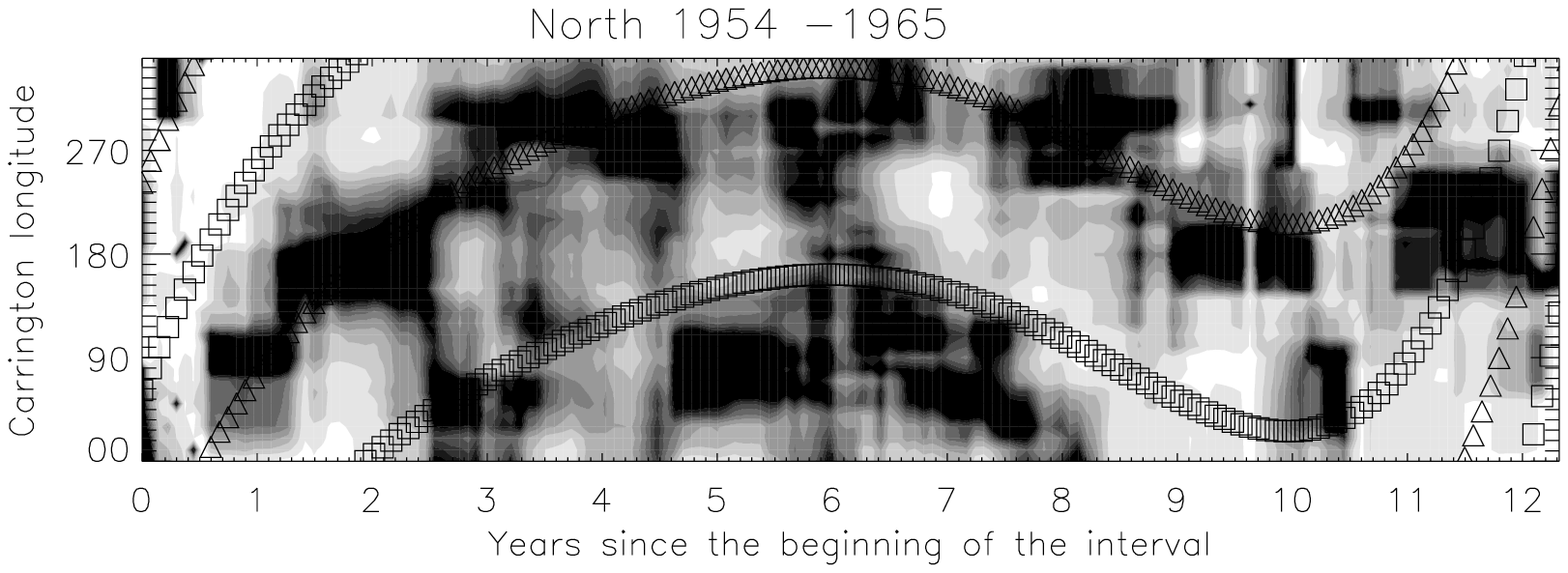,width=8.5 cm}}
\centerline{\epsfig{file=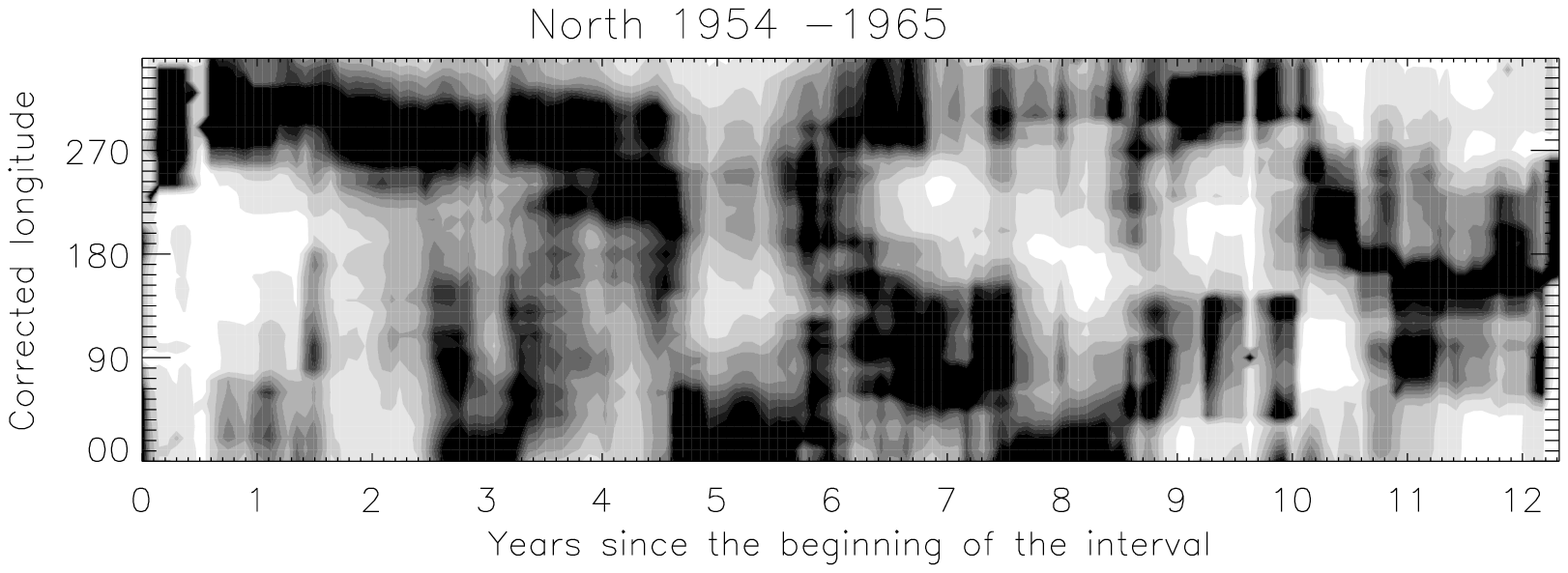,width=8.5 cm}}
\caption{The longitudinal distribution of  the sunspot area during cycle No. 19.
 The vertical axis denotes the longitude and horizontal axis the time.
The upper panel shows the observed Carrington longitudes
and the expected migration path of the two active longitudes
(shown by squares and triangles) given by Eq.~(\ref{Eq:migr})
with $B=3.40$ and $\Omega_0=14.33$ deg$\,$day$^{-1}$.
The lower panel shows the same plot but after the longitude correction, i.e.
 subtraction of the migration path.
For better visualization, each plot was filtered using the Lee
 filter (\cite{lee86} 1986).
}
\label{Fig:No19}
\end{figure}

\begin{figure}
\centerline{\epsfig{file=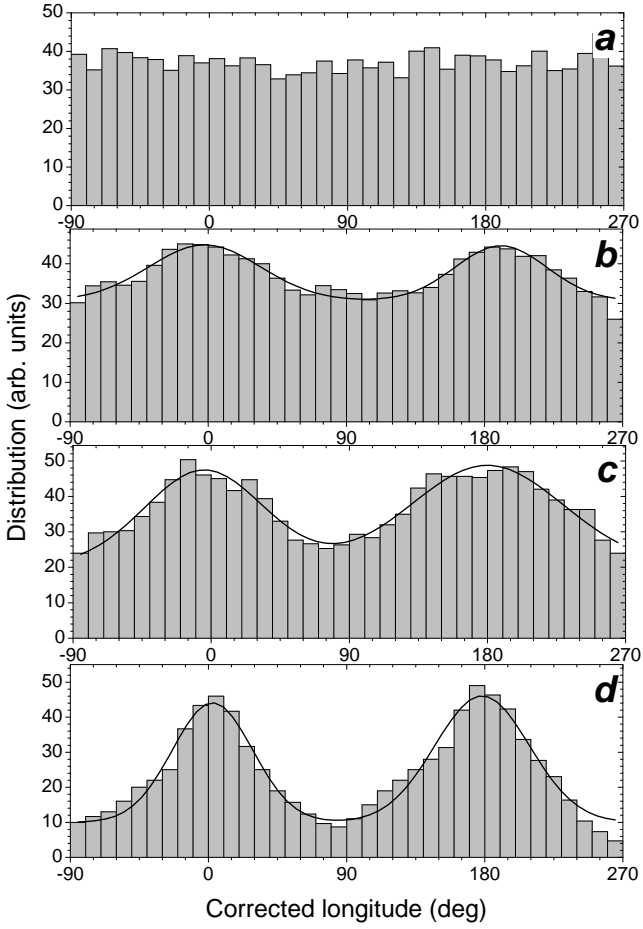,width=8.5 cm}}
\caption{Longitudinal distributions of sunspot occurrence in the Northern
 hemisphere for the period 1878--1996.
 a) Actual sunspots (area weighted) in the Carrington frame.
 The distribution is nearly isotropic.
 b) The same as panel a) but in the dynamic reference frame,
  the non-axisymmetry measure is $\Gamma=0.11$.
 c) Only the position of one dominant spot for each Carrington rotation
is considered, $\Gamma=0.19$.
 d) The same as in panel c) but for 6-months averages, $\Gamma=0.43$.
The solid line depicts the best fit double Gaussian.
 }
\label{Fig:dist_N}
\end{figure}
\begin{figure}
\centerline{\epsfig{file=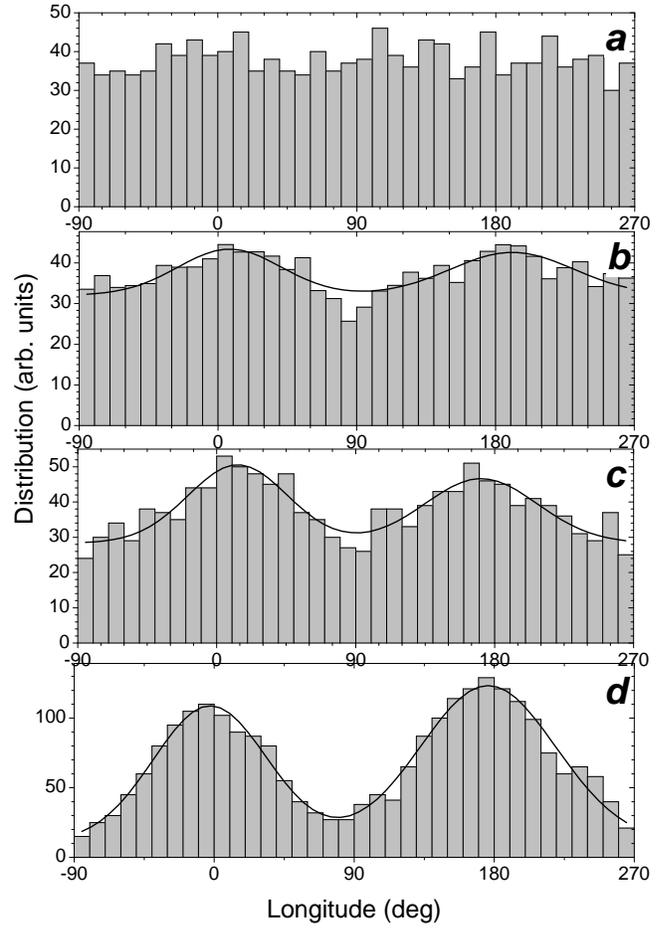,width=8.5 cm}}
\caption{The same as Fig.~\ref{Fig:dist_N} but for the Southern hemisphere.
The value of the non-axisymmetry $\Gamma$ is 0.09, 0.15 and 0.39
 for the panels b), c) and d), respectively.
 }
\label{Fig:dist_S}
\end{figure}
The model predicts that for the time $T_{\rm C}$ the Carrington system
makes one full rotation, while the new reference frame does
$\Omega_j/\Omega_{\rm C}$ rotations.
The difference in longitude accumulated between the two systems over many Carrington periods
 results in migration of the active longitudes, including
 a number of full rotations (see BU03).
In the following, we discard the full rotations and consider only the circular
 longitude.
In the new reference frame, the longitude of a $k$-th spot
 in the $i$-th Carrington rotation, $\tilde\lambda_{ki}$,
 is found as follows:
\begin{equation}
\tilde\lambda_{ki}=\lambda_{ki}-\Lambda_i-n\,360^\circ,
\label{Eq:corr}
\end{equation}
where $\lambda_{ki}$ is the longitude in the Carrington frame, and
$n$ is defined to keep $\tilde\lambda_{ki}$ within the range [$0^\circ$;360$^\circ$].
The transformation to the new reference system is illustrated in
 Fig.~\ref{Fig:No19}, where the upper panel shows the observed
 longitude-vs-time pattern of sunspot occurrence during the solar
 cycle 19.
The gray-scale of the plot represents the normalized area
 of spots occupying a given longitude.
We note that the cycloid-like shape of the active longitude is
 a typical signature of the differential rotation.
The normalized area of a $k$-th sunspot in the $i$-th Carrington rotation is defined as
\begin{equation}
A_{ki}=S_{ki}/\sum_j{S_{ji}},
\label{Eq:A}
\end{equation}
where $S$ is the observed area of the spot corrected for the projection
 effect, and the sum is taken over all spots in the given Carrington rotation.
We consider the normalized area in order to suppress   the dependence of
the spot area on the phase of the  solar cycle.
Otherwise, the analysis would be dominated by large spot groups
 around cycle maxima, with the loss of information about active longitudes during
 the times of minima.
The expected migration path of the active longitude, calculated
according to our model is shown in the upper panel of Fig.~\ref{Fig:No19}
 by squares.
Triangles denote the same active longitude but shifted by $\pm 180^\circ$.
One can see that the expected migration path follows the observed concentration
 of sunspots.
Accordingly, if we correct the observed Carrington longitudes of sunspots
 for the expected migration, the active longitudes should be constant in
the new coordinate frame.
This is shown in the lower panel of Fig.~\ref{Fig:No19}.
The sunspot  area is clustered around the two active longitudes at
 about 90$^\circ$ and $270^\circ$.
One active longitude is generally more active than the other at a given time,
and the dominant activity switches between the two active longitudes,
 indicating a flip-flop phenomenon (cf. BU03).

The  separation of the active longitudes in the dynamic reference
 frame is clearly seen in Fig.~\ref{Fig:dist_N} which shows  histograms
 of the area weighted sunspot occurrence in the corrected longitude
  for the whole data set in the Northern hemisphere.
The sunspot distribution in the Carrington frame shows no
 preferred longitudes (Fig.~\ref{Fig:dist_N}a).
The same distribution, but in the dynamic system, shows a clear
 preference to cluster at two corrected longitudes (Fig.~\ref{Fig:dist_N}b).
Note that this histogram shows the longitude distribution of actual
 spot group areas in the dynamic frame (Eq.~\ref{Eq:corr}), without
any filtering, smoothing or other processing of the raw data.
This is consistent with our earlier result (BU03) that the signature
 of the migrating active longitudes is totally smeared out in the Carrington
 system within 1-2 solar cycles, while a careful account for the migration
 of the active longitudes allows us to reveal their persistence.
A more pronounced double-peaked distribution of the sunspot
area is obtained when considering not all spots but only {\it the largest}
spot within each Carrington rotation (Fig.~\ref{Fig:dist_N}c).
Because of the flip-flop effect the two active longitudes are clearly revealed.
We  emphasize that further processing and averaging of the data significantly
 increase the revealed non-axisymmetry because of the suppression of the
 axisymmetric part.
For instance, semiannual averaging produces a very significant double-peaked
distribution (Fig.~\ref{Fig:dist_N}d and Fig.~5 in BU03).
Very similar histograms are obtained for the Southern hemisphere
 (Fig.~\ref{Fig:dist_S}).
The filtered distributions (panels c and d) are shown here only for the
 purpose of visualisation, and further we will deal only with the raw data
 distribution shown in Figs.~\ref{Fig:dist_N}b and ~\ref{Fig:dist_S}b.


\section{Differential rotation parameters}


Let us now estimate the parameters of the differential rotation
 (Eq.~\ref{Eq:Omegai}) which fit the raw sunspot group data.
For this purpose we employ the least mean squares method.
First, we define the deviation between the model and the data as
\begin{equation}
\epsilon_{ki}={\rm min}\left(\tilde\lambda_{ki}\,;
 \tilde\lambda_{ki}-180^{\circ}\right),
\end{equation}
where indices $i$ and $k$ denote the Carrington rotation number
 and the spot group within this rotation, respectively.
The total discrepancy is then defined as
\begin{equation}
{\cal E}={1\over N}\sum_{i} \sum_{k} A_{ki}\ \epsilon^2_{ki},
\end{equation}
where $N$=$\sum_{i} \sum_{k} A_{ki}$ is the number of Carrington
 rotations with observed sunspots.
Varying the values of $B$ and $\Omega_0$ in Eq.~(\ref{Eq:Omegai}) we
search for a pair which minimizes the discrepancy ${\cal E}$.
Note that we also vary the value of $\Lambda_0$ to obtain the
 minimum discrepancy.

\begin{figure}
\centerline{\epsfig{file=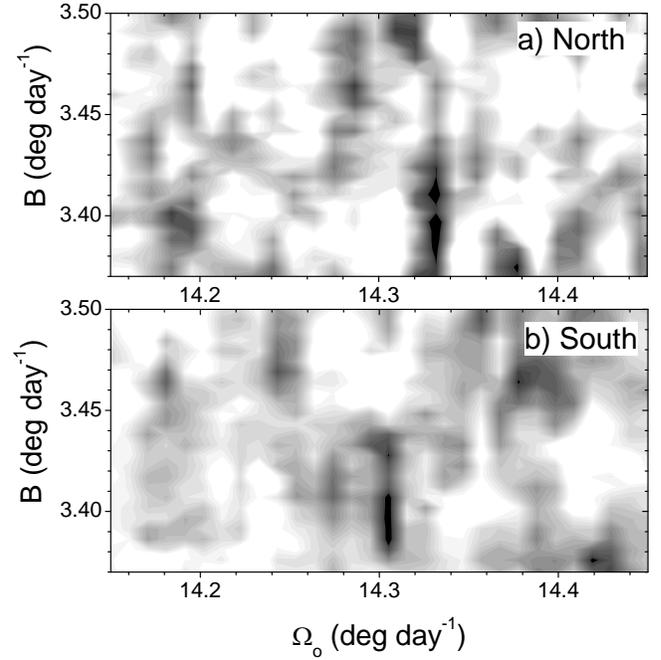,width=8.5 cm}}
\caption{The distribution of the $\chi^2$ values in
 the space of the model parameters $B$ and sidereal $\Omega_0$
(Eq.~\ref{Eq:Omegai}) for the Northern and Southern hemispheres.
The black area depicts the 68\% confidence region ($\Delta\chi^2=2.3$).
 }
\label{Fig:B_O}
\end{figure}

In order to evaluate the confidence intervals for the best-fitting model parameters,
 one needs to estimate the uncertainties of the observed data with respect
 to the fitting model.
From the best-fitting parameters corresponding to the ${\cal E}_{\rm min}$ we can
 build the distribution of the raw data around the expected model
(Fig.~\ref{Fig:dist_N}b).
This distribution is nearly double Gaussian with the standard deviation
 of about $\sigma=70^{\circ}$.
We adopt this value as a measure of the random scattering of the data with respect to the
 model.
Then we can compute an analogue of $\chi^2$ for our model as follows,
\begin{equation}
\chi^2={\cal E}/\sigma^2.
\end{equation}
Note that this is only a scaling of ${\cal E}$, i.e. the best-fitting
 parameters remain the same.
The value of $\chi^2$ is about 0.5 per formal degree of freedom implying that,
 although the model fits the data quite well, the data are slightly autocorrelated
 (i.e. the effective number of degrees of freedom is only half the number
 of points).
The $\chi^2$ distribution in the space of the model parameters $B$ and
 sidereal $\Omega_0$
 is shown in Fig.~\ref{Fig:B_O}.

The best-fitting parameters are
 $B=3.40\pm 0.03$\,deg day$^{-1}$ and
 $\Omega_0=14.33\pm 0.01$\,deg day$^{-1}$ for the Northern hemisphere, and
 $B=3.39\pm0.02$\,deg day$^{-1}$ and
 $\Omega_0=14.31\pm 0.01$\,deg day$^{-1}$ for the Southern hemisphere,
where the quoted errors correspond to the 90\% confidence interval for
one parameter of interest with $\Delta\chi^2=\chi^2-{\rm min}(\chi^2)=2.71$.
The results for the North and South being close to each other
 are nevertheless somewhat different.
The difference is formally significant but, taking the uncertainties of
 the parameter values obtained above as a rough estimate, we can only
 mention its indicative nature.
The distributions in Figs.~\ref{Fig:dist_N} and \ref{Fig:dist_S}
 in the dynamic reference frame were built
 using the above best-fitting values of the model parameters.
The average value of $B$ is close to that obtained by BU03.
We note that the result from BU03 corresponds to a local
 minimum near $\Omega_0\approx$14.2 and $B=3.46$ deg day$^{-1}
 $in Fig.~\ref{Fig:B_O}.

\begin{figure}
\centerline{\epsfig{file=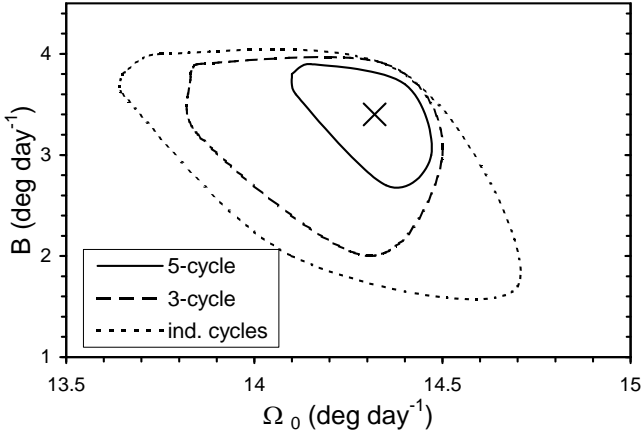,width=8.5 cm}}
\caption{The area of parameters of the differential rotation in the Northern hemisphere
 defined using the entire studied interval of 1878-1996 (large cross)
 as well as from sub-intervals: individual cycles (dotted), 3-cycle intervals
 (dashed) and 5-cycle intervals (solid).
 }
\label{Fig:cycles}
\end{figure}

When repeating the same procedure   for individual cycles,
 we obtained the best-fitting parameters $B$ varying
 from 1.5 to 4 deg day$^{-1}$, and $\Omega_0$ from 13.7 to 14.7 deg day$^{-1}$
 (see Fig.~\ref{Fig:cycles}).
There is a general tendency that smaller $B$ are paired with
 larger $\Omega_0$.
Despite the large spread of the best-fitting parameters for individual cycles,
 the necessity for the differential rotation is apparent,
 because no good fit can be found for $B=0$.
 Formal averaging over individual cycles yields the values of
 $B=3.1\pm 0.6$ deg day$^{-1}$ and $\Omega_0=14.2\pm 0.24$ deg day$^{-1}$.
The analysis for individual cycles could not possibly be used to check the
 persistence of the active longitudes, because each cycle is fitted independently.
However, extending the studied interval not only systematically tightens the
 allowed area in the parameter space  towards the small area determined for
 the entire data set, as illustrated in Fig.~\ref{Fig:cycles},
  but also proves the persistency of the phenomenon.

\begin{figure}
\centerline{\epsfig{file=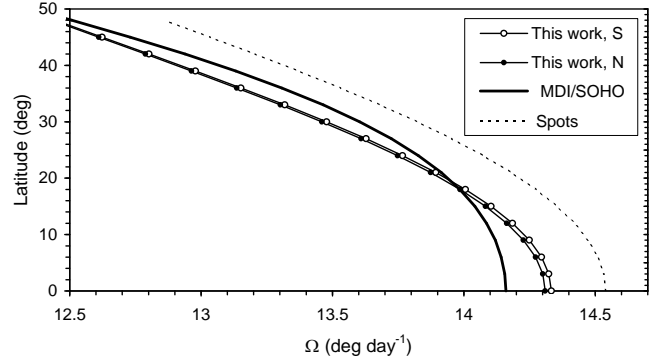,width=8.5 cm}}
\caption{The sidereal differential rotation of the active longitudes
 determined in this work compared with that obtained using
 surface Doppler shifts (SOHO/MDI, \cite{shou98} 1998) and sunspots
 (\cite{balt86} 1986).
 }
\label{Fig:diff}
\end{figure}

In Fig.~\ref{Fig:diff} we compare the present results for the differential
rotation of the active longitudes with measurements obtained
from sunspots (\cite{balt86} 1986) and surface plasma Doppler shifts
(SOHO/MDI, \cite{shou98} 1998).
Note that the SOHO/MDI data were originally approximated by
Eq.~(\ref{Eq:Omega}).
As seen from Fig.~\ref{Fig:diff}, at all latitudes the active longitudes rotate
significantly slower than individual sunspots, which is in agreement with
the previous finding by BU03. This implies that the active longitudes
are not linked to the depth where developed
sunspots are anchored.
The difference  in the rotation of the active longitudes and
 individual spots is about 0.2\,deg\,day$^{-1}$ which produces a lag of about 2 full
 rotations during a solar cycle as reported by BU03.
A comparison with the SOHO/MDI measurements is more difficult because of the
different shapes of the used approximations. It appears that the active longitudes
 rotate faster than the local plasma at low latitudes ($\psi<18^\circ$)
 and vice-versa at higher latitudes.


\section{A measure of the non-axisymmetry}


The existence of the active longitudes implies a long-term asymmetry in
the sunspot longitudinal distribution which is related to a non-axisymmetric component
 of the dynamo mechanism.
In order to quantify this we introduce a measure of the non-axisymmetry in the following way.
Let us choose the value of $\Lambda_0$ so that the two active longitudes correspond to the
 corrected longitudes of 0$^\circ$ and 180$^\circ$ (see Fig.~\ref{Fig:dist_N}b).
Depending on the value of the corrected longitude $\tilde\lambda_{ki}$,
 the sunspot with the normalized area $A_{ki}$ contributes to either of the two numbers
as follows.
\begin{eqnarray}
N_1&=&\sum_{k,i}{A_{ki}\,\mbox{, if }|\tilde\lambda_{ki}|<45\degr\mbox{ or
}|\tilde\lambda_{ki}-180\degr|<45\degr},\nonumber\\
N_2&=&\sum_{k,i}{A_{ki}}\,\mbox{, otherwise,}\hskip 3.7cm
\end{eqnarray}
where the summation is taken over all spots in all Carrington rotations.
 Thus, $N_1$ and $N_2$ represent the number of spots, weighted by
 their area, which are close to and far from the active longitudes,
 respectively.
The non-axisymmetry is then defined as
\begin{equation}
\Gamma=\frac{N_1-N_2}{N_1+N_2}.
\label{Eq:gam}
\end{equation}
It can take values from 0 to 1, where 0 corresponds to the
 axisymmetric sunspot distribution and 1 to the non-axisymmetric distribution.

\begin{figure}
\centerline{\epsfig{file=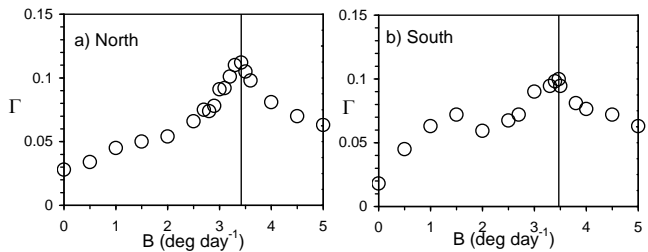,width=8.5 cm}}
\caption{The measure of the non-axisymmetry $\Gamma$ as a function of the
 differential rotation parameter $B$ for the Northern (left panel)
 and Southern (right panel) hemispheres.
 The values of $B=3.42$ and $3.43\ {\rm deg\ day}^{-1}$
  as denoted by the vertical lines, correspond to the maximum
  $\Gamma$ in the Northern and Southern hemispheres, respectively.}
\label{Fig:M_B}
\end{figure}

The non-axisymmetry of the area-weighted sunspot distribution for the entire studied
 series shown in Figs.~\ref{Fig:dist_N}b and \ref{Fig:dist_S}b is
  0.11  and 0.09 for the Northern and Southern hemispheres, respectively.
Without normalizing the spot area (Eq.~\ref{Eq:A}), the distributions
 indicate the non-axisymmetry of 0.07.
We conclude, therefore, that the strength of the nonaxisymmetric component is roughly 1:10 of that
of the axisymmetric one as observed in the sunspot distribution for the last 120 years.
For individual cycles $\Gamma$ takes values from 0.07 to 0.3.
We note that at the best-fitting parameters minimizing discrepancy
${\cal E}$ (see Fig.~\ref{Fig:B_O}),
$\Gamma$ does not necessarily reach the maximum value, being, however, rather close to it.
The dependence of $\Gamma$ on the parameter $B$ is
 shown in Fig.~\ref{Fig:M_B} (parameter $\Omega_0$ is now chosen to provide
 the maximum possible $\Gamma$).
One can see that the relation has a single peak at about $B=3.42$--3.43 deg day$^{-1}$
 (close to the best-fitting value $B=3.39$--3.40 deg day$^{-1}$ minimizing $\cal E$) and decreases when deviating from it.
If the longitudes are not corrected for the migration (i.e., $B=0$),
 the non-axisymmetry is $\Gamma=$0.02--0.03 (see Fig.~\ref{Fig:M_B}a) which is consistent with the
null-hypothesis of the axisymmetric distribution (see below).
This shows again that neglecting the differential rotation results
 in complete smearing of the pattern.

\begin{figure}
\centerline{\epsfig{file=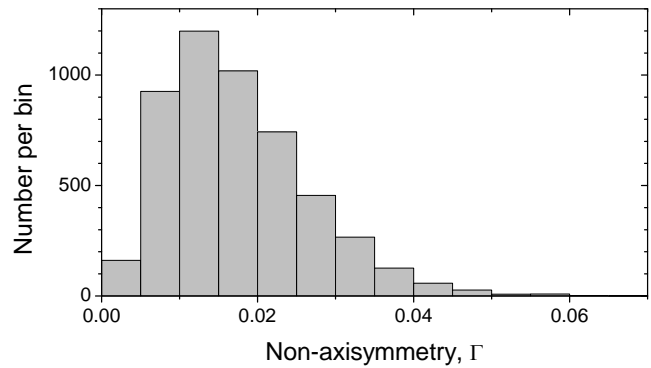,width=8.5 cm}}
\caption{The distribution of the non-axisymmetry $\Gamma$ calculated for
5,000 sets of random-phase sunspot occurrence.
}
\label{Fig:M_dist}
\end{figure}

In order to check the significance of the obtained non-axisymmetry
 measure $\Gamma$ we have performed the following test.
For each Carrington rotation we have randomly permuted all the sunspots,
 i.e., a new random Carrington longitude has been ascribed to each actually observed
 sunspot while keeping its area.
Then the value of $\Gamma$ has been calculated as described above.
We have computed the non-axisymmetry $\Gamma$ for 5,000 sets of such random-phase
 sunspot occurrence.
The distribution of the value of $\Gamma$ shown in
 Fig.~\ref{Fig:M_dist} depicts a nearly Poisson distribution.
According to this simulation, the hypothesis of rotation of
 the active longitudes with a fixed rate
(giving $\Gamma=0.02-0.03$; see Figs.~\ref{Fig:dist_N}a, \ref{Fig:dist_S}a, and
\ref{Fig:M_B}a,b) cannot be distinguished from
 the null hypothesis of the axisymmetric sunspot distribution.
On the other hand,
¨the probability to obtain $\Gamma=0.11$ (0.09)
 for an axisymmetric distribution is $<10^{-6}$
 ($\approx 10^{-5}$) for the Northern (Southern) hemispheres.
This means that the non-axisymmetric component does really exist in the raw data of sunspot
 occurrence  and at the very high significance level.
For the case when only the dominant spot is considered
 (Figs.~\ref{Fig:dist_N}c and \ref{Fig:dist_S}c) $\Gamma=0.15-0.19$,
 implying that nearly 60\% of
 the major spots appear in the vicinity of the active longitudes.
Moreover, the non-axisymmetric mode dominates the semiannually averaged sunspot
 occurrence (Figs.~\ref{Fig:dist_N}d and \ref{Fig:dist_S}d).


\section{Conclusions}


Active longitudes on the Sun were  commonly expected to rotate
 with a constant rate.
This  a priori assumption led previous  researchers to the
 conclusion   that the active longitudes, if exist, are not stable
 in their appearance.
 The recent analysis of the sunspot distribution revealed however that
 there are two persistent active longitudes which migrate
 according to the differential rotation law (BU03).
In this paper we have confirm this  finding and determined the parameters of the
 differential rotation law affecting the active longitude migration.
We emphasize that in the present paper we  employ a different approach
 compared to that used in BU03,
 where   the data were analysed to reveal the underlying regularities.
Here we fit a theoretical model of the differential rotation to the raw data
 without any pre-processing or filtering of the latter.
Previously (BU03) we investigated only the non-axisymmetric part of the sunspot
 distribution while efficiently suppressing the axisymmetric one.
This was criticized by \cite{pelt04} (2005) who claimed that the active longitude pattern
 found in BU03  is an artefact of the used method.
The present analysis, which is based on the raw observed sunspot areas,
 answers their criticism and confirms that the phenomenon of the
 persistent active longitudes is real.

The main conclusions obtained  in the present paper are listed below.

\begin{itemize}

\item
The raw sunspot occurrence data for the last 120 years confirm the
 existence of two preferred
 longitudes 180$^{\circ}$ apart, which migrate in any fixed rotation frame
 but are persistent throughout the entire period studied in the dynamic frame
 defined by Eq.~(\ref{Eq:corr}).

\item
The differential rotation of the active longitudes is well approximated
 by Eq.~(\ref{Eq:Omegai}) with $B=3.40\pm 0.03$ ($3.39\pm 0.02$) deg day$^{-1}$
 and sidereal $\Omega_0=14.33\pm 0.01$ (14.31$\pm 0.01$) deg day$^{-1}$
 for Northern (Southern) hemispheres, respectively.
This is significantly different from the differential rotation of individual spots.
This implies that the depth at which sunspots are formed (and affected
 by the non-axisymmetric component of the field) is different from
 that where developed sunspots are anchored.

\item
The measure of the non-axisymmetry $\Gamma$ (Eq.~\ref{Eq:gam})
 for the raw data is found to be 0.11 (0.09) for the Northern (Southern)
 hemispheres.
These correspond to the significance $<10^{-6}$ ($10^{-5}$), respectively.
The strength of the non-axisymmetric dynamo component is roughly 1/10 of
 that of the axisymmetric one.
Appropriate smoothing, filtering or processing of the data suppresses
 the axisymmetric component and increases the formal measure of the non-axisymmetry.

\end{itemize}

The above findings  imply the existence of a weak, but persistent,
non-axisymetric dynamo component with phase migration.
The relative strength of this mode and the law of its rotation
 obtained here provide new constraints for the development of solar dynamo models
as was recently undertaken by, e.g., \cite{moss99} (1999, 2004, 2005).

\begin{acknowledgement}
We thank Dmitry Sokoloff and David Moss for stimulating and useful discussions.
Solar data have been obtained from the GRO and USAF/NOAA
web site\\ http://science.nasa.gov/ssl/pad/solar/greenwch.htm.
The Academy of Finland is acknowledged for the support, grant 43039.
\end{acknowledgement}

\end{document}